\documentclass[12pt,nofootinbib,superscriptaddress]{revtex4}

\usepackage{amssymb,color,paralist,amsmath,epsfig}
\usepackage{graphicx}
\usepackage{amsfonts}
\usepackage{nicefrac,esint}
\usepackage{verbatim}
\usepackage{float}

\usepackage{relsize}

\usepackage[colorlinks,citecolor=blue,linktoc=all,linkcolor=cyan]{hyperref}
\usepackage[section]{placeins}

\newcommand{\hp}{{\frac{1}{2}}}
\newcommand{\hm}{{-\frac{1}{2}}}

\newcommand{\beq}{\begin{equation}}
\newcommand{\eeq}{\end{equation}}

\newcommand{\ber}{\begin{eqnarray}} 
\newcommand{\eer}{\end{eqnarray}}

\usepackage{color}

\usepackage[normalem]{ulem}

\renewcommand\sout{\bgroup \color[rgb]{0,0.00,1.} \ULdepth=-.5ex \ULset}

\begin{document}

\title{Two-photon exchange correction to the hyperfine splitting in muonic hydrogen}
\author{Oleksandr Tomalak}
\affiliation{Institut f\"ur Kernphysik and PRISMA Cluster of Excellence, Johannes Gutenberg Universit\"at, Mainz, Germany}

\date{\today}

\begin{abstract}
We reevaluate the Zemach, recoil and polarizability corrections to the hyperfine splitting in muonic hydrogen expressing them through the low-energy proton structure constants and obtain the precise values of the Zemach radius and two-photon exchange (TPE) contribution. The uncertainty of TPE correction to S energy levels in muonic hydrogen of 105 ppm exceeds the ppm accuracy level of the forthcoming 1S hyperfine splitting measurements at PSI, J-PARC and RIKEN-RAL.

\end{abstract}

\maketitle

\tableofcontents

\newpage
\section{Introduction}

The first spectroscopy measurements with muonic atoms by the CREMA Collaboration at PSI \cite{Pohl:2010zza} allowed us to study the proton electromagnetic structure with unprecedented precision. The accurate extraction of the proton charge radius from the muonic hydrogen Lamb shift \cite{Pohl:2010zza,Antognini:1900ns} gave the discrepancy to measurements with electrons \cite{Bernauer:2010wm,Bernauer:2013tpr,Mohr:2012tt}; see \cite{Antognini:1900ns,Carlson:2015jba} for recent reviews. This problem is known as the {\it proton radius puzzle}.

The precise spectroscopy measurements require an improvement in the theoretical knowledge of the radiative corrections. The dominant theoretical uncertainty in the proton size extractions from the Lamb shift is coming from the graph with two exchanged photons. Thus this proton structure correction triggered a lot of attention in the theoretical community \cite{Pachucki:1996zza,Faustov:1999ga,Faustov:1999gax,Pineda:2002as,Pineda:2004mx,Nevado:2007dd,Carlson:2011zd,Hill:2012rh,Miller:2012ne,Birse:2012eb,Alarcon:2013cba,Gorchtein:2013yga,Peset:2014jxa,Tomalak:2015hva,Caprini:2016wvy,Hill:2016bjv}. The dispersive estimates of the two-photon exchange (TPE) contribution give $\Delta E_{\mathrm{TPE}} (\mu \mathrm{H})  = 33.2 (2.0)~\mu \mathrm{eV}$ \cite{Carlson:2011zd, Birse:2012eb, Antognini:2013jkc}, which is far below the observed discrepancy in $ 310~\mu \mathrm{eV}$. However, the uncertainty of this contribution is comparable with the experimental accuracy in $ 2~\mu \mathrm{eV}$. Additionally, such TPE estimates depend on the model of the subtraction function in the forward Compton scattering, which is an active research field last years \cite{Carlson:2011zd,Miller:2012ne,Hill:2012rh,Birse:2012eb,Alarcon:2013cba,Gorchtein:2013yga,Peset:2014jxa,Tomalak:2015hva,Caprini:2016wvy,Hill:2016bjv}.

The new highly precise insights on the proton electromagnetic structure will be obtained by the forthcoming measurements of 1S hyperfine splitting (HFS) in muonic hydrogen with an unprecedented ppm precision by the CREMA~\cite{Pohl:2016tqq} and FAMU~\cite{Adamczak:2016pdb,Dupays:2003zz} Collaborations as well as at J-PARC~\cite{Ma:2016etb}. In these experiments, the expected accuracy level is two orders of magnitude smaller than the theoretical knowledge of the TPE correction with 213 ppm uncertainty in the dispersive estimate \cite{Carlson:2011af} and 109 ppm in the effective field theory approach \cite{Peset:2016wjq}. The leading TPE effects of the proton structure in HFS are expressed in terms of the proton spin structure functions and form factors \cite{Zemach:1956zz,Iddings:1959zz,Iddings:1965zz,Drell:1966kk,Faustov:1966,Faustov:1970,Bodwin:1987mj,Faustov:2001pn,Martynenko:2004bt,Carlson:2008ke,Carlson:2011af,Hagelstein:2015egb,Peset:2016wjq,Tomalak:2017owk}. Consequently, the dominant uncertainty from the TPE correction can be reduced by the precise measurements of the proton electric and magnetic form factors in the low-$Q^2$ region \cite{Denig:2016dqo} and studies of the proton spin structure functions $ g_1$ and $g_2 $ by EG4, SANE and g2p experiments at JLab \cite{Zheng:2009zza,SANE:2011aa,Zielinski:2017gwp}.

In Ref. \cite{Tomalak:2017owk}, we proved the standard expressions for the TPE correction \cite{Carlson:2008ke,Carlson:2011af} expressing it in terms of the forward lepton-proton scattering amplitudes. With the aim to decrease the uncertainty of the $\alpha^5$ TPE contribution to HFS, we reevaluate the Zemach, recoil and polarizability corrections expressing the region with small photon virtuality in terms of proton radii \cite{Karshenboim:2014maa,Karshenboim:2014vea}, which was introduced by Karshenboim to constrain the values of the electric and magnetic radii from the atomic spectroscopy measurements, and moments of the spin structure functions. We exploit the elastic proton form factors fit, which is based on the unpolarized and polarization transfer world data \cite{Bernauer:2010wm,Bernauer:2013tpr}, and the latest parametrization of the proton spin structure functions \cite{Kuhn:2008sy,griffioen,Sato:2016tuz,Fersch:2017qrq}. Additionally, we express the polarizability correction in terms of the measurable spin asymmetry, which provides a direct relation to the experimental observables.

The paper is organized as follows. We describe the standard framework of the TPE correction to S-level HFS and evaluate the proton and inelastic intermediate states contributions in Sect. \ref{hfs_correction}. Afterwards, we present the comparison with previous computations. We give our conclusions with an outlook of the forthcoming 1S HFS measurements in Sect. \ref{conclusions}.

\section{Two-photon exchange correction to the hyperfine splitting}
\label{hfs_correction}

The two-photon exchange (TPE) contribution to the nS-level hyperfine splitting (HFS) $ \delta E^{\mathrm{HFS}}_{\mathrm{nS}} $ is expressed in terms of the relative correction $ \Delta_{\mathrm{HFS}} $ and the leading order nS-level HFS $ E^{\mathrm{HFS},0}_{\mathrm{nS}} $ (Fermi energy) as \footnote{Note that the muon anomalous magnetic moment contribution should be treated separately \cite{Eides:2000xc}.}
\ber
 \delta E^{\mathrm{HFS}}_{\mathrm{nS}} & = & \Delta_{\mathrm{HFS}} E^{\mathrm{HFS},0}_{\mathrm{nS}} , \\
 E^{\mathrm{HFS},0}_{\mathrm{nS}} & = & \frac{8}{3} \frac{ m_r^3 \alpha^4}{ M m} \frac{\mu_P}{n^3},
\eer
where $M$ and $m$  are the proton and the lepton masses, $ m_r = M m / ( M + m )$ is the reduced mass, $ \mu_P \approx 2.793 $ is the proton magnetic moment and $ \alpha \approx 1/137 $ is the electromagnetic coupling constant.

The TPE correction is given by a sum of diagrams with proton and with inelastic intermediate states. Conventionally, it is expressed as a sum of the Zemach correction $ \Delta_{\mathrm{Z}} $, the recoil correction $ \Delta^{\mathrm{p}}_{\mathrm{R}} $ and the polarizability correction $ \Delta^{\mathrm{pol}} $ \cite{Eides:2000xc}:
\ber \label{traditional}
\Delta_{\mathrm{HFS}} & = & \Delta_{\mathrm{Z}} + \Delta^{\mathrm{p}}_{\mathrm{R}} +  \Delta^{\mathrm{pol}} , \\
\label{Zemach_correction}
\Delta_{\mathrm{Z}}& = &\frac{8 \alpha m_r}{\pi \mu_P}  \int \limits^{\infty}_{0} \frac{\mathrm{d} Q}{Q^2} \left( G_M\left(Q^2\right)  G_E\left(Q^2\right)  - \mu_P \right),
\eer
\ber
\Delta^{\mathrm{p}}_{\mathrm{R}}  & = &  \frac{\alpha}{\pi \mu_P} \int \limits^{\infty}_{0} \frac{\mathrm{d} Q^2}{Q^2} \left\{ \frac{  \left [ 2 + \rho\left(\tau_l\right) \rho\left(\tau_P \right) \right ]  F_D\left(Q^2\right) + 3 \rho\left(\tau_l\right) \rho \left(\tau_P\right) F_P\left(Q^2\right)   }{ \sqrt{\tau_P} \sqrt{1+\tau_l} + \sqrt{\tau_l} \sqrt{1+\tau_P} } - \frac{4 m_r}{Q} G_E\left(Q^2\right)  \right\} \nonumber \\ 
&& \qquad \qquad \quad ~~ \times G_M\left(Q^2\right) - \frac{\alpha}{\pi \mu_P} \frac{m}{M} \int \limits^{\infty}_{0} \frac{\mathrm{d} Q}{Q} \rho (\tau_l) \left( \rho (\tau_l) - 4 \right) F_P^2 \left(Q^2\right),  \label{recoil_correction} \\
\Delta^{\mathrm{pol}}  & = &  \frac{2 \alpha}{\pi \mu_P} \int \limits^{\infty}_{0} \frac{\mathrm{d} Q^2}{Q^2} \int \limits^{\infty}_{\nu^{\mathrm{inel}}_{\mathrm{thr}}} \frac{\mathrm{d} \nu_\gamma}{\nu_\gamma}  \frac{\left [ 2 + \rho\left(\tau_l\right) \rho\left(\tilde{\tau}\right) \right ]  g_1 \left(\nu_\gamma, Q^2 \right)  -  3 \rho\left(\tau_l\right) \rho\left(\tilde{\tau} \right)  g_2 \left(\nu_\gamma, Q^2 \right)   / \tilde{\tau}}{  \sqrt{\tilde{\tau}} \sqrt{1+\tau_l} + \sqrt{\tau_l} \sqrt{1+\tilde{\tau}}   } \nonumber \\
&& + \frac{\alpha}{\pi \mu_P} \frac{m}{M} \int \limits^{\infty}_{0} \frac{\mathrm{d} Q}{Q} \rho (\tau_l) \left( \rho (\tau_l) - 4 \right) F^2_P \left(Q^2\right), \label{polar_correction}
\eer
with the photon energy $\nu_\gamma$ and the photon virtuality $Q^2$. $ F_D (Q^2) $, $ F_P(Q^2) $, $ G_E (Q^2) $, $ G_M(Q^2) $ are the Dirac, Pauli, Sachs electric and magnetic proton form factors (FFs), $g_1 \left(\nu_\gamma, Q^2 \right)$ and $g_2 \left(\nu_\gamma, Q^2 \right)$ are the spin-dependent inelastic proton structure functions. The following definitions were introduced:
\ber
 \tau_l = \frac{Q^2}{4 m^2}, ~~~~~ \tau_P = \frac{Q^2}{4 M^2}, ~~~~~  \tilde{\tau} = \frac{\nu_\gamma^2}{Q^2}, ~~~~~\rho(\tau) = \tau - \sqrt{\tau ( 1 + \tau )}.  \label{taus}
\eer
 The inelastic threshold is given by $ \nu^{\mathrm{inel}}_{\mathrm{thr}} = m_\pi +  \left( m_\pi^2 + Q^2 \right) / \left( 2 M \right)$, with the pion mass $ m_{\pi} $.
 
In the following sections, we evaluate the contributions of Eqs. (\ref{Zemach_correction})-(\ref{polar_correction}) separately performing the low-energy expansion in the region of low photon virtuality.

\subsection{Zemach and recoil correction evaluation}

The Zemach correction can be evaluated accounting for the measured values of the proton charge and magnetic radii. We split the $Q$-integration in the Zemach contribution at the small enough scale $ Q_0 $ and exploit the radii expansion at low $ Q^2 $ \cite{Karshenboim:2014vea}, thus
\ber \label{Zemach_expansion}
\Delta_{\mathrm{Z}}& = & \frac{4 \alpha m_r Q_0 }{3\pi} \left( - r^2_E- r^2_M + \frac{r^2_Er^2_M}{18} Q^2_0\right) \nonumber \\
&+&  \frac{8 \alpha m_r}{\pi}  \int \limits^{\infty}_{Q_0} \frac{\mathrm{d} Q}{Q^2} \left( \frac{G_M\left(Q^2\right)  G_E\left(Q^2\right)}{\mu_P}  - 1 \right) ,
\eer
with the approximate value $ Q_0 \sim 0.1-0.2~\mathrm{GeV}$ and the definition of the proton radii:

\ber
r^2_{E(M)} = - \frac{6}{G_{E(M)} \left(0 \right)} \left. \frac{\mathrm{d} G_{E(M)} \left(Q^2 \right)}{ \mathrm{d} Q^2}  \right \vert_{Q^2 = 0}.
\eer
For the numerical evaluations, we exploit the elastic proton form factors fit of Ref. \cite{Bernauer:2013tpr}, which is based on a global analysis of the electron-proton scattering data at $ Q^2 < 10~\mathrm{GeV}^2$ with an account of TPE corrections. The resulting uncertainty is evaluated as a sum of the form factors uncertainties \cite{Bernauer:2010wm,Bernauer:2013tpr} for $Q^2 > Q^2_0$ and radii uncertainties for $Q^2 < Q^2_0$ in quadrature. We add the point $Q^2 = 0$ with a zero uncertainty to the fit of form factors. We select $Q_0 = 0.15 ~\mathrm{GeV} $ in the following and estimate the error due to this choice as a difference between our results with $ Q_0 = 0.15 ~\mathrm{GeV} $ and $ Q_0 = 0.2 ~\mathrm{GeV} $. We account for the $Q^4$ and $Q^6$ terms, exploiting the chiral perturbation theory expansion coefficients \cite{Horbatsch:2016ilr}, and we add the uncertainty of the higher-order contributions as the difference between the calculation with higher-order terms in expansion and result based on Eq. (\ref{Zemach_expansion}), which contributes $13~\mathrm{ppm}$ to the Zemach correction. We substitute the values of the electric charge radius $ r^e_E =  0.879 \pm 0.008 ~\mathrm{fm}$ from the electron-proton scattering data \cite{Bernauer:2013tpr} as well as $ r^{\mu \mathrm{H}}_E =  0.84087 \pm 0.00039 ~\mathrm{fm}$ from the muonic hydrogen spectroscopy experiments \cite{Antognini:1900ns}. For the proton magnetic radius, we choose the extraction of the A1 Collaboration $ r^e_M =  0.799 \pm 0.017 ~\mathrm{fm}$ \cite{Bernauer:2013tpr} and the later more conservative analysis of Ref. \cite{Arrington:2015ria} $ r^W_M =  0.844 \pm 0.038 ~\mathrm{fm}$. In the following, we study the systematic uncertainty due to the pure knowledge of the proton radii performing the calculation for all possible combinations of chosen electric and magnetic radii.

\begin{figure}[H]
\begin{center}
\includegraphics[width=0.71\textwidth]{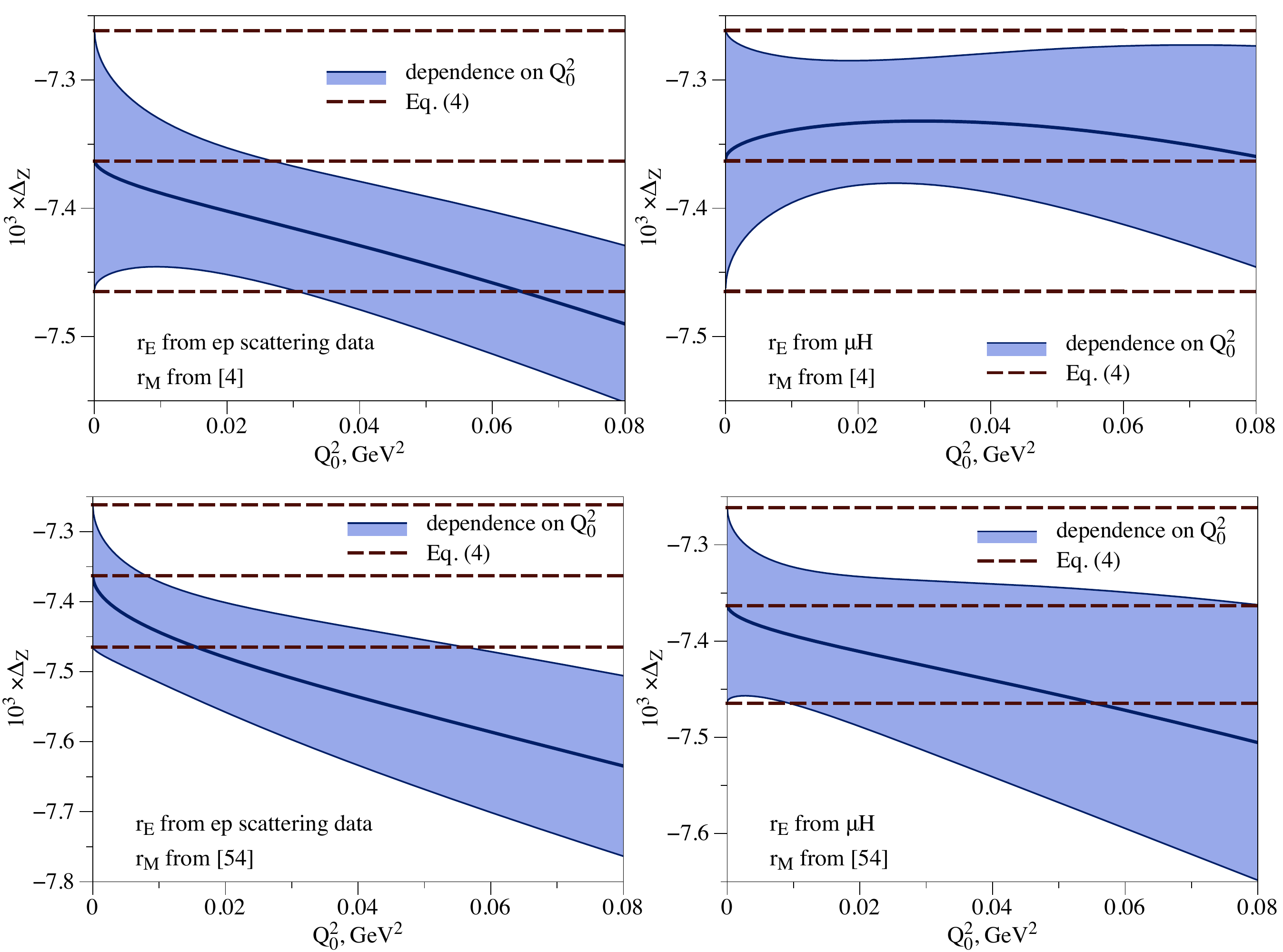}
\end{center}
\caption{Consistency check: the dependence of the Zemach correction $\Delta_\mathrm{Z}$ on the splitting parameter $Q_0$ in Eq. (\ref{Zemach_expansion}). Left panel: charge radius value from the electron-proton scattering data. Right panel: muonic hydrogen charge radius value. Upper panel: magnetic radius of Ref. \cite{Bernauer:2013tpr}. Lower panel: magnetic radius of Ref. \cite{Arrington:2015ria}. }
\label{consist}
\end{figure}

As a consistency check, we show the dependence of the Zemach contribution on the splitting parameter $Q_0$ in Fig. \ref{consist}. The upper plots with the magnetic radius value of A1 Collaboration are closer to the plateau behavior at small $Q_0$, which has to appear for the consistent experimental input. However, neither $\mu \mathrm{H}$ nor the electron-proton scattering charge radius passes this check. 

In Table \ref{Zemach_detail} we provide results for different contributions to Zemach term with the corresponding uncertainties. In the calculation with the magnetic radius $r^e_M$ of Ref. \cite{Bernauer:2013tpr}, the main uncertainty comes from the error in the proton magnetic radius and form factors in the region of A1/MAMI data $Q^2 \lesssim (0.6-1)~\mathrm{GeV}^2$. The dependence on the splitting parameter is larger in the case of the electron-proton scattering charge radius corresponding to a better consistency of plots in the right panel of Fig. \ref{consist}. In the calculation with the magnetic radius $r^W_M$ of Ref. \cite{Arrington:2015ria}, the uncertainty is dominated by the conservative error estimate of the proton magnetic radius extraction. The error due to the choice of the splitting parameter is also enhanced in the case of larger $r_M$, which can be read off from the lower panel of Fig. \ref{consist} representing pure consistency of $r^W_M$ with other electromagnetic proton properties. 
\begin{table}[H] 
\begin{center}
\begin{tabular}{|c|c|c|c|c|c|}
\hline
 $\Delta_Z $ (ppm) & $r^e_E $, $r^e_M $  &   $r^{\mu \mathrm{H}}_E $, $r^e_M $   &   $r^e_E $, $r^\mathrm{W}_M $  &   $r^{\mu \mathrm{H}}_E $, $r^\mathrm{W}_M $    \\ \hline
 $ r^2_E + r^2_M - \frac{r^2_Er^2_M}{18} Q^2_0 $ &   $-$1581(34) &   $-$1508(30) & $-$1662(73) &  $-$1590(71) \\ \hline
$r_E$ uncertainty & 16 & 1 & 16 & 1 \\ \hline
$r_M$ uncertainty  & 30 & 30 & 71 & 71 \\ \hline
Uncertainty in choice of $Q_0$  & 23 & 1 & 49  & 26 \\ \hline
Higher-order expansion terms & \multicolumn{4}{c|}{13(13)}   \\ \hline
FFs, $  Q^2_0 < Q^2 < 0.6~\mathrm{GeV}^2 $ &   \multicolumn{4}{c|}{5890(34)}   \\ \hline
FFs above $ 0.15~\mathrm{GeV}^2  $  &  \multicolumn{4}{c|}{647(5) }   \\ \hline
FFs above $ 0.6~\mathrm{GeV}^2  $  &  \multicolumn{4}{c|}{36(1) }   \\ \hline
FFs above $ 1~\mathrm{GeV}^2  $  &  \multicolumn{4}{c|}{8.2(0.5) }   \\ \hline
Zemach correction  &   $-$7406(56) &   $-$7333(48) & $-$7487(95) &  $-$7415(84) \\ \hline 
\end{tabular}
\caption{Contributions to Zemach term with corresponding uncertainties.} \label{Zemach_detail}
\end{center}
\end{table}

The evaluation with the form factor parametrizations at high-$Q^2$ of Refs. \cite{Arrington:2007ux,Venkat:2010by} gives the same $36~\mathrm{ppm}$ as the fit of A1 Collaboration from the region $Q^2 > 0.6~\mathrm{GeV}^2 $. However, the earlier parametrization of Ref. \cite{Kelly:2004hm} results in $39~\mathrm{ppm}$. The possible $ 3~\mathrm{ppm} $ error is negligible in the evaluation of the resulting uncertainty in quadrature.

The difference between two results based on $\mu$H spectroscopy and electron data in $73~\mathrm{ppm}$ can give a hint on the correct radius value in new HFS measurements with ppm accuracy level.  Only an improved precision of the magnetic form factor and radius as well as the reduction of the uncertainty in the polarizability correction will make it possible. We also evaluate the recoil correction $ \Delta^{\mathrm{p}}_{\mathrm{R}}$ and the sum $ \Delta_\mathrm{Z} + \Delta^{\mathrm{p}}_{\mathrm{R}} $ performing the similar radii expansion and present the results in Table \ref{HFS_elastic} of Sect. \ref{comparison}. We evaluate the error of $ \Delta_\mathrm{Z} + \Delta^{\mathrm{p}}_{\mathrm{R}} $ adding uncertainties from the form factors under the integral in quadrature.

For completeness, we provide a detailed study of the recoil correction in Table \ref{recoil_detail}. The main contribution and uncertainty come from form factors in the region of A1/MAMI data $Q^2 \lesssim (0.6-1)~\mathrm{GeV}^2$. Other form factor parametrizations at high-$Q^2$ of Refs. \cite{Arrington:2007ux,Venkat:2010by,Kelly:2004hm} contribute only around 1 ppm from the region $Q^2 > 0.6~\mathrm{GeV}^2 $ within the uncertainty of the estimate in Table \ref{recoil_detail}. The recoil correction in the calculation with the larger value of the magnetic radius $r^W_M$ is 2 ppm below the result with $r^e_M$ and shows a 2 ppm dependence on the choice of the splitting parameter.
\begin{table}[H] 
\begin{center}
\begin{tabular}{|c|c|c|c|c|c|}
\hline
 $\Delta^\mathrm{p}_\mathrm{R} $ (ppm) & $r^e_E $, $r^e_M $  &   $r^{\mu \mathrm{H}}_E $, $r^e_M $   &   $r^e_E $, $r^\mathrm{W}_M $  &   $r^{\mu \mathrm{H}}_E $, $r^\mathrm{W}_M $    \\ \hline
Form factors expansion &   621.61(0.05) &   621.33(0.05)& 619.15(0.26) &  618.86(0.25) \\ \hline
$r_E$ uncertainty & 0.01 &  0.0003  & 0.01 & 0.0003 \\ \hline
$r_M$ uncertainty  & 0.05 &  0.05 & 0.25& 0.25 \\ \hline
Uncertainty in choice of $Q_0$  & 0.1 & 0.5 & 2.2 & 2.6 \\ \hline
Higher-order expansion terms & \multicolumn{4}{c|}{0.3(0.3)} \\ \hline
FFs, $  Q^2_0 < Q^2 < 0.6~\mathrm{GeV}^2 $  &   \multicolumn{4}{c|}{223.2(5.1)}   \\ \hline
FFs above $ 0.15~\mathrm{GeV}^2  $  &  \multicolumn{4}{c|}{24.9(2.5) }   \\ \hline
Form factors above $ 0.6~\mathrm{GeV}^2  $  & \multicolumn{4}{c|}{1.5(1)}   \\ \hline
Form factors above $ 1~\mathrm{GeV}^2  $  &  \multicolumn{4}{c|}{0.5(0.4)}  \\ \hline
Recoil correction  &   846.6(6.2) & 846.4(6.2) & 844.2(6.6) & 843.9(6.7) \\ \hline
\end{tabular}
\caption{Contributions to recoil correction with corresponding uncertainties.} \label{recoil_detail}
\end{center}
\end{table}

Additionally, we obtain the precise value for the Zemach radius $ r_Z $ which is defined as
\ber \label{radius}
 r_Z = - \frac{\Delta_Z}{2 \alpha m_r }.
\eer
Substituting the electric charge radius from the scattering data and the magnetic radius $r^e_M$ ($r^W_M$), we evaluate the Zemach radius as $r_Z =  1.0544 \pm 0.0079  ~\mathrm{fm} $ ($r_Z =  1.0660 \pm 0.0135  ~\mathrm{fm} $). With the substitution of the charge radius from the muonic hydrogen spectroscopy and the magnetic radius $r^e_M$ ($r^W_M$), the Zemach radius is given by $r_Z = 1.0440 \pm 0.0068 ~\mathrm{fm} $ ($r_Z =  1.0557 \pm 0.0120  ~\mathrm{fm} $). The results are in reasonable agreement between each other and with the extractions from atomic spectroscopy of Refs. \cite{Antognini:1900ns,Dupays:2003zz,Volotka:2004zu,Hagelstein:2017cbl,Dorokhov:2017nzk} as well as previous evaluations of Eqs. (\ref{Zemach_correction}, \ref{radius}) \cite{Friar:2003zg,Carlson:2008ke,Distler:2010zq}.

\subsection{Polarizability correction evaluation}
\label{polarizability_HFS}

For the numerical evaluation of the polarizability correction, we subtract the leading moment of the spin structure function $ g_1 $ and separate contributions from the structure functions $ g_1 $ and $ g_2 $ \cite{Hagelstein:2015egb}:
\ber
\Delta^{\mathrm{pol}}_{\mathrm{0}}  & = & \Delta^{\mathrm{pol}}_{\mathrm{1}}  + \Delta^{\mathrm{pol}}_{\mathrm{2}}, \label{g1_HFS} \\
\Delta^{\mathrm{pol}}_{\mathrm{1}} & = &  \int \limits^{\infty}_{0}  \mathrm{I}_{\mathrm{I}_1} (Q) \mathrm{d} Q  +  \int \limits^{\infty}_{0}  \mathrm{I}_{\mathrm{g}_1} (Q) \mathrm{d} Q,  \\
\Delta^{\mathrm{pol}}_{\mathrm{2}} & = & \int \limits^{\infty}_{0}  \mathrm{I}_{\mathrm{g}_2} (Q) \mathrm{d} Q , \label{g2_HFS}
\eer
with the corresponding integrands:
\ber
 \mathrm{I}_{\mathrm{I}_1} (Q)& = & \frac{\alpha}{\pi \mu_P} \frac{m}{M} \frac{\rho (\tau_l) \left( \rho (\tau_l) - 4 \right)}{Q}  \left \{ 4 I_1 \left( Q^2 \right) + F^2_P \left(Q^2\right) \right \},  \\
\mathrm{I}_{\mathrm{g}_1} (Q)& = & \frac{4 \alpha}{\pi \mu_P} \int \limits^{\infty}_{\nu^{\mathrm{inel}}_{\mathrm{thr}}} \frac{\mathrm{d} \nu_\gamma}{Q \nu_\gamma} \left ( \frac{ 2  +    \rho(\tau_l) \rho(\tilde{\tau})      }{  \sqrt{\tilde{\tau}} \sqrt{1+\tau_l} + \sqrt{\tau_l} \sqrt{1+\tilde{\tau}}   } -  \frac{m \rho (\tau_l) \left( \rho (\tau_l) - 4 \right)}{\nu_\gamma} \right ) g_1 \left(\nu_\gamma, Q^2 \right),\nonumber \\ \label{g1int}\\
 \mathrm{I}_{\mathrm{g}_2} (Q)& = & -\frac{12 \alpha}{\pi \mu_P}\int \limits^{\infty}_{\nu^{\mathrm{inel}}_{\mathrm{thr}}} \frac{\mathrm{d} \nu_\gamma}{Q \nu_\gamma \tilde{\tau}} \frac{    \rho(\tau_l) \rho(\tilde{\tau}) g_2 \left(\nu_\gamma, Q^2 \right) }{  \sqrt{\tilde{\tau}} \sqrt{1+\tau_l} + \sqrt{\tau_l} \sqrt{1+\tilde{\tau}}}, \label{g2int}
\eer
where the first moment $I_1 \left( Q^2 \right) $ of the structure function $ g_1 $ is given by
\ber
I_1 \left( Q^2 \right) & = & \int \limits^{\infty}_{ \nu^{\mathrm{inel}}_{\mathrm{thr}}} g_1 \left(\nu_\gamma, ~Q^2\right) \frac{M \mathrm{d} \nu_\gamma}{\nu_\gamma^2}, \qquad \quad I_1(0) = - \frac{\left(\mu_P - 1 \right)^2}{4}. \label{I1_moment}
\eer

\begin{figure}[H]
\begin{center}
\includegraphics[width=0.54\textwidth]{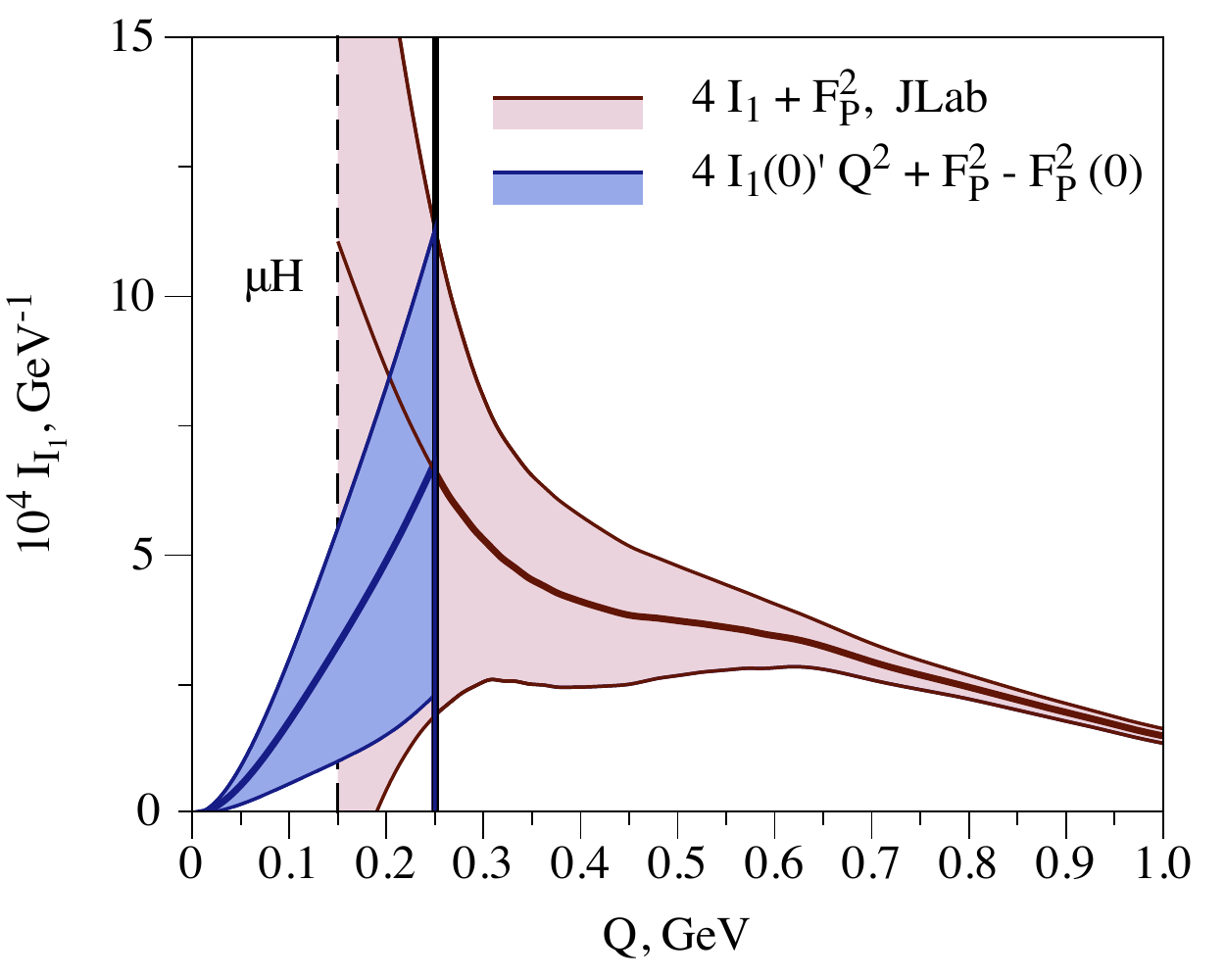}
\end{center}
\caption{JLab HFS integrand $ \mathrm{I}_{\mathrm{I}_1}$ connected to the low-$Q^2$ behavior.}
\label{with_one_subtraction2}
\end{figure}

In order to evaluate the contribution from $ 4 I_1 + F^2_P $, we approximate $ I_1 \left(Q^2\right) = I_1\left(0\right) + I_1(0)' Q^2 $ up to $ Q_{\mathrm{I_1}} = 0.25 ~\mathrm{GeV} $ with  the low-energy constant $ I_1(0)' = 7.6 \pm 2.5 ~\mathrm{GeV}^{-2} $ \cite{Prok:2008ev}. For larger $Q^2$, we exploit the spin structure functions data parametrization of Refs. \cite{Kuhn:2008sy,griffioen,Sato:2016tuz,Fersch:2017qrq} (JLab parametrization). We show the corresponding $ Q^2$-dependence of the integrand $ \mathrm{I}_{\mathrm{I}_1}$ in Fig. \ref{with_one_subtraction2}. 

The low-$Q^2$ and the larger-$Q^2$ integrands have an intersection point slightly above the lowest data at $Q^2 \sim 0.05~\mathrm{GeV}^2$, which was used in the structure functions parametrization \cite{Kuhn:2008sy,griffioen,Sato:2016tuz,Fersch:2017qrq}. At smaller values of $Q^2$, the uncertainty of the data parametrization rapidly increases, and the integration can give an overestimated value. The recent JLAB data \cite{Zielinski:2017gwp} confirms the smaller effective value of $ I_1(0)'$ for the parametrization. Consequently, the criterion of the same integrand values and similar uncertainties in both regions helps us to choose $Q_{\mathrm{I_1}}$. We estimate the inaccuracy due to this choice as a difference in the HFS correction between calculations with two splitting parameters: $  Q_{\mathrm{I_1}} = 0.2 ~\mathrm{GeV}  $ and $  Q_{\mathrm{I_1}} = 0.25 ~\mathrm{GeV} $, which evaluates to 8.6 ppm, and add it in quadrature. 

\begin{figure}[H]
\begin{center}
\includegraphics[width=0.48\textwidth]{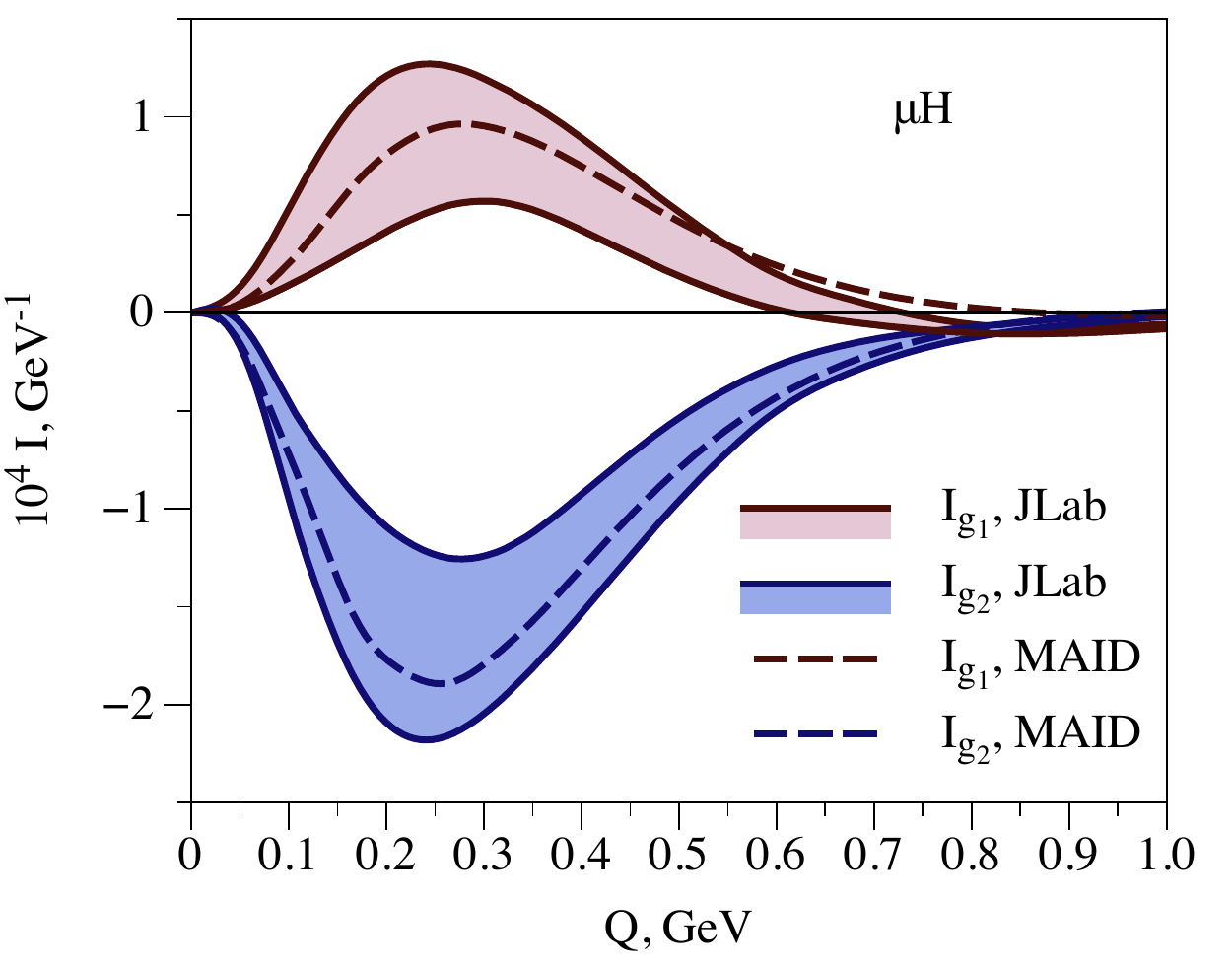}
\end{center}
\caption{MAID and JLab integrands $ \mathrm{I}_{\mathrm{g}_1}$ and $ \mathrm{I}_{\mathrm{g}_2}$.}
\label{g1_g22}
\end{figure}

For the remaining polarizability corrections $\Delta_1^{\mathrm{pol}}$ and $\Delta_2^{\mathrm{pol}}$ from the proton spin structure functions we use the JLab parametrization only, which is in fair agreement with the MAID model \cite{Drechsel:2000ct,Drechsel:2007if} in the region of low $ Q^2 $, see Fig. \ref{g1_g22} for details.

We add the uncertainties coming from the Pauli form factor $ F_P $ \cite{Bernauer:2010wm,Bernauer:2013tpr}, the spin structure functions $ g_1,~g_2 $ and the parameter $ I_1(0)' $ in quadrature under the HFS integrand and treat the uncertainties from the two $ Q $-integration regions in $ \Delta_1 $ and $ \mathrm{I}_{\mathrm{I}_1} $ contributions as uncorrelated uncertainties. One of important error sources in the resulting polarizability correction is $ 47~\mathrm{ppm}$ uncertainty from the error of $ I_1(0)'$.

Additionally, we add the $23~\mathrm{ppm}$ error of the higher-order terms in the low-energy expansion as a difference of our evaluation and the calculation \cite{Tomalak:2017owk} with the replacement of the following leading moments of the spin structure functions at $Q^2 < 0.25~\mathrm{GeV}^2$:
\ber
I_2 \left( Q^2 \right) & = & \frac{2M^2}{Q^2} \int \limits^{x^\mathrm{inel}_\mathrm{thr}}_0 g_2 \left( x_{\mathrm{Bj}}, ~Q^2\right) \mathrm{d} x_{\mathrm{Bj}} = \frac{1}{4} F_P \left( Q^2 \right) G_M \left( Q^2 \right), \label{I2_moment} \\
I^{(3)}_1 \left( Q^2 \right) & = & \frac{8M^4}{Q^4} \int \limits^{x^\mathrm{inel}_\mathrm{thr}}_0 x_{\mathrm{Bj}}^2 g_1 \left( x_{\mathrm{Bj}}, ~Q^2\right) \mathrm{d} x_{\mathrm{Bj}} \underset{Q^2 \to 0}{\longrightarrow} \frac{Q^2 M^2}{2 \alpha} \gamma_0,  \\
I^{(3)}_2 \left( Q^2 \right) & = & \frac{8M^4}{Q^4} \int \limits^{x^\mathrm{inel}_\mathrm{thr}}_0 x_{\mathrm{Bj}}^2 g_2 \left( x_{\mathrm{Bj}}, ~Q^2\right) \mathrm{d} x_{\mathrm{Bj}} \underset{Q^2 \to 0}{\longrightarrow} \frac{Q^2 M^2}{2 \alpha} \left( \delta_{\mathrm{LT}} - \gamma_0 \right), \label{I32_moment}
\eer
 by the low-energy constants \cite{Drechsel:2000ct,Drechsel:2007if,Drechsel:2002ar,Prok:2008ev,Pascalutsa:2014zna,Lensky:2017dlc}:
\ber
 \delta_{\mathrm{LT}} & = & \left( 1.34 \pm 0.17 \right) \times 10^{-4} ~\mathrm{fm}^4, \\
  \gamma_0 & = & \left( -1.01 \pm 0.13\right) \times 10^{-4} ~\mathrm{fm}^4.
\eer

We present the results for different contributions of Eqs. (\ref{g1_HFS})-(\ref{g2_HFS}) to the S-level HFS in $\mu \mathrm{H}$ and compare them to Refs. \cite{Faustov:2001pn,Carlson:2008ke,Hagelstein:2017cbl} in Table \ref{muH_pol}, where for results of Ref. \cite{Faustov:2001pn} we have accounted for the convention conversion correction of Ref. \cite{Carlson:2011af}. Though the contributions from the structure functions $ g_1 $ and $ g_2 $ are slightly different to previous dispersive evaluations of Refs. \cite{Faustov:2001pn,Carlson:2011af,Carlson:2008ke}, the resulting polarizability correction is in good agreement with the results of Ref. \cite{Faustov:2001pn}: $ \Delta_0^{\mathrm{pol}} = 410 \pm 80 ~\mathrm{ppm} $ and Ref. \cite{Carlson:2008ke}: $ \Delta_0^{\mathrm{pol}} = 351 \pm 114 ~\mathrm{ppm}$. All dispersive evaluations are in contradiction to the chiral perturbation theory result \cite{Hagelstein:2017cbl} due to the large difference in the $ \mathrm{I}_{\mathrm{I}_1} $ contribution.
\begin{table}[H] 
\begin{center}
\begin{tabular}{|c|c|c|c|c|c|c|}
\noalign{\vskip 5mm}    
\hline
$ \Delta $, ppm & $ \mathrm{I}_{\mathrm{I}_1}$  & $ \mathrm{I}_{\mathrm{g}_1}$ &  $ \mathrm{I}_{\mathrm{I}_1} + \mathrm{I}_{\mathrm{g}_1}  $  & $ \mathrm{I}_{\mathrm{g}_2}$ & $ \Delta^{\mathrm{pol}}_{0} $    \\ \hline
this work  & $ 402(91) $ & $ 27(15) $ & $ 429(84)  $  &  $ -65(20) $ & $ 364(89) $  \\ \hline
Hagelstein et al. \cite{Hagelstein:2017cbl} & $-$21  & 58 & $ 37  $  &  $ -98 $ & $ -61^{+70}_{-52} $  \\ \hline
Carlson et al. \cite{Carlson:2008ke} &  &  & $ 370(112)  $  &  $ -19(19) $ & $ 351(114) $  \\ \hline
Martynenko et al. \cite{Faustov:2001pn} &  &  & $ 468 $  &  $ -58$ & $ 410(80) $  \\ \hline
\end{tabular}
\caption{TPE correction to the S level HFS in $ \mu \mathrm{H}$.}  \label{muH_pol}
\end{center}
\end{table}

Within the dispersion relation approach of Ref. \cite{Tomalak:2017owk}, we express the polarizability correction $ \Delta^{\mathrm{pol}}_{\mathrm{0}} $ directly in terms of the measurable inclusive inelastic $ lp$ cross sections as
\ber
\Delta^{\mathrm{pol}}_{\mathrm{0}} & = &  \frac{3 M m }{\pi  e^2 \mu_P}  \int \limits^{ \infty}_{\omega_{\mathrm{thr}}}  \frac{\sigma^{\mathrm{inel}}_{\hp \hp}\left(\omega^\prime \right) - \sigma^{\mathrm{inel}}_{\hp \hm}\left(\omega^\prime \right) }{\sqrt{\omega^{\prime 2}-m^2}}  \mathrm{d} \omega^\prime + \frac{\alpha}{\pi \mu_P} \frac{m}{M} \int \limits^{\infty}_{0} \frac{\mathrm{d} Q}{Q} \beta_1 \left(\tau_l\right) F^2_P \left(Q^2\right),
\eer
where $ \sigma^{\mathrm{inel}}_{h \lambda} $ denotes the inclusive inelastic cross section with the incoming lepton (proton) helicity $ h $($ \lambda $). The integration starts from the inelastic  threshold, i.e. the pion production threshold $ \omega_{\mathrm{thr}} = m + m_\pi (2 M+ 2 m + m_\pi)/ (2 M) $. Such an expression can be used for the independent direct evaluation of the polarizability correction from the data input.

\subsection{Comparison with literature}
\label{comparison}
\begin{figure}[H]
\begin{center}
\includegraphics[width=0.59\textwidth]{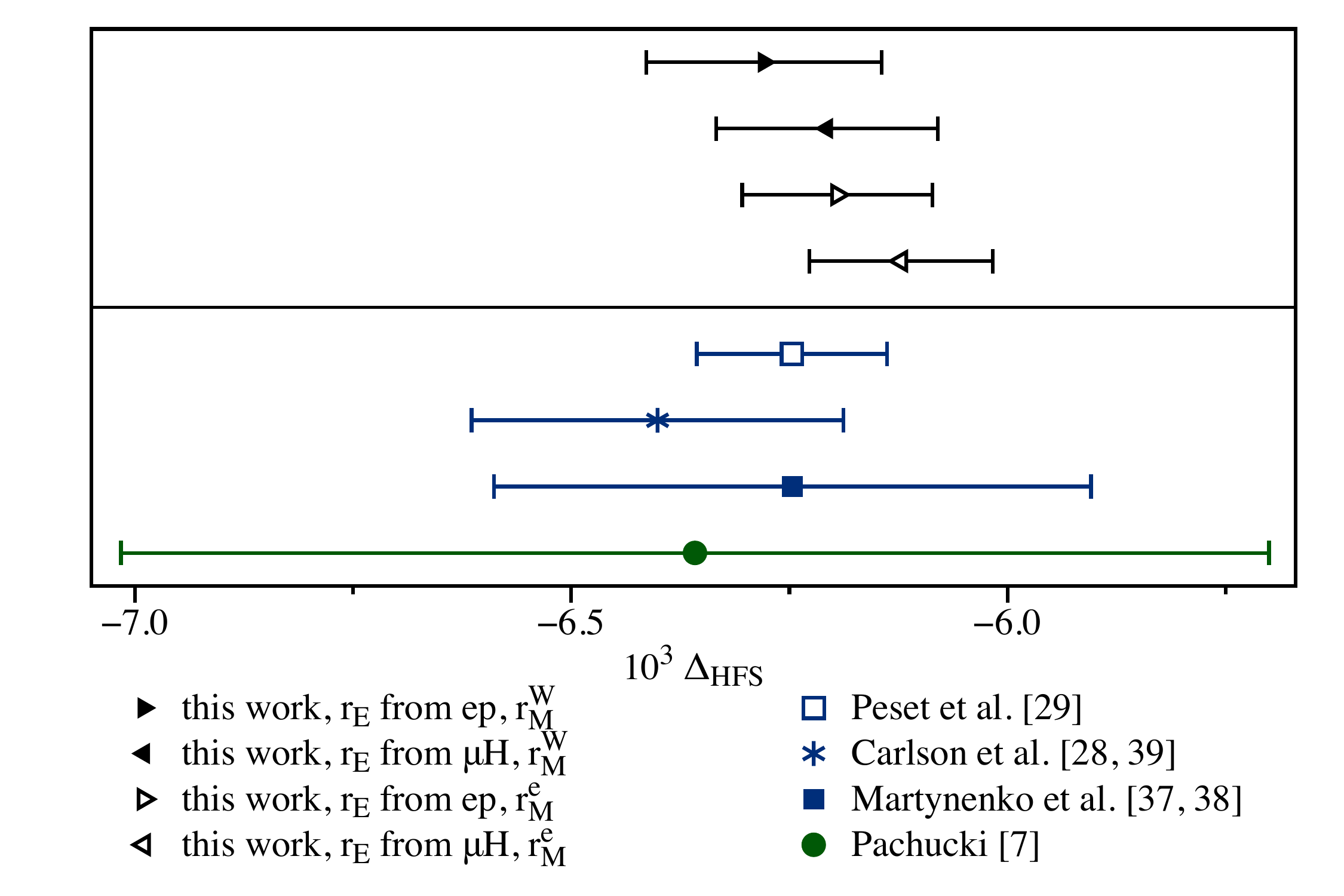}
\end{center}
\caption{Two-photon exchange correction to the S-level hyperfine splitting in $ \mu \mathrm{H} $. Results are presented in the chronological order starting from below.}
\label{comparison2}
\end{figure}

In Table \ref{HFS_elastic} \footnote{The errors of $ \Delta_{\mathrm{Z}} $, $ \Delta^{\mathrm{p}}_{\mathrm{R}}  $, $ \Delta^{\mathrm{pol}}_{0}  $, $ \Delta_{\mathrm{Z}} + \Delta^{\mathrm{p}}_{\mathrm{R}} $ and $ \Delta_{\mathrm{HFS}} $ are strongly correlated. Therefore, we evaluate them separately.} we compare our results for different HFS contributions to the previous evaluations of Refs. \cite{Pachucki:1996zza,Faustov:2001pn,Martynenko:2004bt,Carlson:2008ke,Carlson:2011af,Peset:2016wjq,Hagelstein:2015egb,Hagelstein:2017cbl}, where we have subtracted the recoil correction of order $ \alpha^2$ \cite{Carlson:2008ke,Carlson:2011af}, the radiative correction to the Zemach contribution \cite{Carlson:2008ke,Carlson:2011af,Peset:2016wjq} and the convention conversion correction of Ref. \cite{Carlson:2011af} when it is needed \cite{Faustov:2001pn,Martynenko:2004bt}. The absolute value of the Zemach contribution is smaller than results of previous estimates \cite{Friar:2003zg,Carlson:2008ke,Carlson:2011af} based on the existing form factors parametrizations before the A1/MAMI data, which has a larger value of the magnetic form factor at low-$Q^2$ region. The recoil correction is in reasonable agreement with other estimates \cite{Carlson:2011af,Hagelstein:2017cbl}. The polarizability correction is in good agreement with dispersive calculations, though all dispersive results are in contradiction to the ChPT prediction.
\begin{table}[H] 
\begin{center}
\begin{tabular}{|c|c|c|c|c|c|}
\hline
$ \Delta $, (ppm) &  $ \Delta_{\mathrm{Z}} $  & $ \Delta^{\mathrm{p}}_{\mathrm{R}}  $ &  $ \Delta_{\mathrm{Z}} + \Delta^{\mathrm{p}}_{\mathrm{R}} $  & $ \Delta^{\mathrm{pol}}_{0}  $ & $ \Delta_{\mathrm{HFS}} $ \\ \hline
this work, $ \mu \mathrm{H}$ $r_E$, $r^W_M $ &    $-$7415(84) & 844(7) & $-$6571(87) & 364(89) & $-$6207(127)  \\ \hline
this work, electron $r_E$, $r^W_M $ &   $-$7487(95) &   844(7)  &  $-$6643(98) & 364(89) & $-$6279(135)   \\ \hline
this work, $ \mu \mathrm{H}$ $r_E$, $r^e_M $ &    $-$7333(48) & 846(6) & $-$6486(49) & 364(89) & $-$6122(105)  \\ \hline
this work, electron $r_E$, $r^e_M $ &   $-$7406(56) &   847(6)  &  $-$6559(57) & 364(89) & $-$6195(109)   \\ \hline
Hagelstein et al. \cite{Hagelstein:2017cbl} & & & & $-61^{+70}_{-52}$ &  \\ \hline
Peset et al. \cite{Peset:2016wjq}    &  &  &  & & $-$6247(109) \\ \hline
Carlson et al. \cite{Carlson:2008ke,Carlson:2011af}  &  $-$7587 & 835 & $-$6752(180) & 351(114) & $-$6401(213)  \\ \hline
Martynenko et al. \cite{Martynenko:2004bt}  & $-$7180 &  & $-$6656 & 410(80) & $-$6246(342)  \\ \hline
Pachucki  \cite{Pachucki:1996zza} &  $-$8024 &  & $-$6358 & 0(658) & $-$6358(658)  \\ \hline
\end{tabular}
\caption{Two-photon exchange contribution to the S-level hyperfine splitting in $ \mu \mathrm{H} $.} \label{HFS_elastic}
\end{center}
\end{table}
We finish the comparison to previous results for the total HFS correction in Fig. \ref{comparison2}. The difference from Refs. \cite{Carlson:2008ke,Carlson:2011af} is mainly due to the smaller value of the Zemach radius in our evaluation. The smaller value of the polarizability contribution in Refs. \cite{Hagelstein:2015egb,Hagelstein:2017cbl} causes the largest discrepancy to our results.

\section{Conclusions and outlook}
\label{conclusions}

In view of the forthcoming high-precision measurements of the 1S hyperfine splitting in muonic hydrogen with $ \mathrm{ppm} $ precision level \cite{Pohl:2016tqq,Ma:2016etb,Adamczak:2016pdb}, we provide the corresponding best estimates of the TPE correction in Table \ref{muH_tab}. The uncertainty of our result is 100 times larger than the expected experimental accuracy. The proton state contribution allows one to determine the precise value of the Zemach radius, which shows a consistent within 1-1.5$\sigma$ dependence on the proton radii. The error of the polarizability contribution is almost two times larger than the uncertainty of the Zemach term in the evaluation with the proton magnetic radius of Ref. \cite{Bernauer:2013tpr} and is of the similar size exploiting the radius of Ref. \cite{Arrington:2015ria}. It is dominated by the pure knowledge of $ I_1(0)' $ and spin structure functions $g_1,~g_2$. The forthcoming data from EG4, SANE and g2p experiments at JLab on the proton spin structure functions $ g_1,~g_2 $ \cite{Zheng:2009zza,SANE:2011aa,Zielinski:2017gwp} will improve the knowledge of the polarizability correction. The precise measurements of the proton magnetic form factors at low $Q^2$ \cite{Denig:2016dqo} and the reextraction of the magnetic radius \cite{Tomalak:2016vbf} will allow us to decrease the uncertainty of the Zemach contribution.
\begin{table}[h] 
\begin{center}
\begin{tabular}{|c|c|c|c|c|c|}
\hline
 $\Delta_{\mathrm{HFS}} $ (ppm) & $r^e_E $, $r^e_M $  &   $r^{\mu \mathrm{H}}_E $, $r^e_M $   &   $r^e_E $, $r^\mathrm{W}_M $  &   $r^{\mu \mathrm{H}}_E $, $r^\mathrm{W}_M $    \\ \hline
Zemach, $ \Delta_{\mathrm{Z}} $ &   $-$7406(56) &   $-$7333(48) & $-$7487(95) &  $-$7415(84) \\ \hline
Recoil, $ \Delta^{\mathrm{p}}_{\mathrm{R}} $  & 846.6(6.2) & 846.4(6.2) & 844.2(6.6) & 843.9(6.7) \\ \hline
Polarizability, $ \Delta^{\mathrm{pol}}_{0}  $  &  364(89) & 364(89) & 364(89) & 364(89)   \\ \hline
Total, $ \Delta_{\mathrm{HFS}} $ &   $-$6195(109) & $-$6122(105) & $-$6279(135) & $-$6207(127)   \\ \hline \hline
Zemach radius, $r_Z$ ($\mathrm{fm} $)  &  1.0544(0.0079)  & 1.0440(0.0068) &  1.0660(0.0135) & 1.0557(0.0120)     \\ \hline
\end{tabular}
\caption{Finite-size TPE contributions to the hyperfine splitting of the S energy levels in $ \mu $H and Zemach radius. Results are shown for values of charge radii from the electron-proton scattering data and $\mu \mathrm{H}$ spectroscopy and two magnetic radius extractions of Refs. \cite{Bernauer:2013tpr,Arrington:2015ria}. } \label{muH_tab}
\end{center}
\end{table}

Consequently, after accounting for all corrections at the $ 1-10~\mathrm{ppm} $ level, the forthcoming measurements can constrain the low-$Q^2$ proton structure contribution to HFS $ \Delta_{\mathrm{structure}} $ with the following combination of the radii and $ I_1(0)' $:
\ber
\Delta_{\mathrm{structure}} = - \frac{4 \alpha }{3\pi} \left ( m_r Q_0 \left(r^2_E + r^2_M \right) +  \frac{m}{M}  \frac{h\left( \tau_l \right)}{\mu_P} I_1(0)' m^2 \right ),
\eer
where
\ber
 h(\tau) =  \left(9 - 4 \tau \right) \tau^2 + \frac{15}{2} \ln \left( \sqrt{\tau} + \sqrt{1+\tau} \right) -\frac{1}{2} \left( 15 + 22 \tau - 8 \tau^2 \right)\sqrt{\tau \left(1+\tau \right)},
\eer
and $ \tau_l $ is taken at the point $ Q =  Q_{\mathrm{I_1}} \sim \left( 0.1 - 0.3\right) \mathrm{GeV}$, up to which we use the low-energy expansion of $ I_1 (Q^2)$.

\section{Acknowledgments}
We thank Marc Vanderhaeghen, Carl Carlson, Randolf Pohl, Vladimir Pascalutsa, Jan Bernauer and Mikhail Gorchtein for useful discussions. We thank Keith Griffioen, Sebastian Kuhn, Nevzat Guler and Jacob Ethier for providing us with results on the proton spin structure functions and Slava Tsaran for his script for an online access of MAID. This work was supported by the Deutsche Forschungsgemeinschaft (DFG) through Collaborative Research Center ``The Low-Energy Frontier of the Standard Model'' (SFB 1044), and Graduate School ``Symmetry Breaking in Fundamental Interactions'' (DFG/GRK 1581).

\end{document}